\documentclass[structabstract]{aa}  

\usepackage{natbib}
\bibpunct{(}{)}{;}{a}{}{,}
\usepackage{graphicx}
\usepackage{txfonts}
\usepackage{longtable}
\usepackage{epstopdf}
\usepackage{subfigure}
\usepackage{multirow}
\begin{document}
   \title{Wave propagation and energy transport in the magnetic network of the Sun}

   \author{G. Vigeesh
          \inst{1,2},
          S. S. Hasan\inst{1}
	  \and
	  O. Steiner\inst{2}
          }

   \institute{Indian Institute of Astrophysics, Block II Koramangala,
              Bangalore-560034, India\\
              \email{vigeesh@iiap.res.in, hasan@iiap.res.in}
         \and
             Kiepenheuer-Institut f\"ur Sonnenphysik, Sch\"oneckstrasse 6, 
             79104 Freiburg, Germany\\
             \email{steiner@kis.uni-freiburg.de}
             }

   \date{Received 8 May 2009 / Accepted 11 September 2009}

\titlerunning{Wave propagation \& energy transport in the magnetic network}
\authorrunning{Vigeesh et al.}
 
  \abstract
   {}
   {We investigate wave propagation and energy transport in magnetic elements,
which are representatives of small scale magnetic flux concentrations in  the
magnetic network on the Sun.
This is a continuation of earlier work by Hasan et al. (2005). The new
features in the present investigation include a quantitative evaluation of
the energy transport in the various modes and for different field strengths,
as well as the effect of the boundary-layer thickness on wave propagation.}
   {We carry out 2-D MHD numerical simulations of magnetic flux concentrations 
for strong and moderate magnetic fields for which $\beta$  (the ratio of gas to 
magnetic pressure) on the tube axis at the photospheric base is 0.4 and 1.7,
respectively. Waves are excited in the tube and ambient medium by a
transverse impulsive motion of the lower boundary.}
{The nature of the modes excited depends on the value of $\beta$. Mode
conversion occurs in the moderate field case when the fast mode crosses the
$\beta=1$ contour. In the strong field case the
fast  mode undergoes conversion from predominantly magnetic to predominantly
acoustic when waves are leaking from the interior of the flux concentration to
the ambient medium.
We also estimate the energy fluxes in the acoustic and magnetic modes and find 
that in the strong field case, the vertically directed acoustic wave fluxes 
reach spatially averaged, temporal maximum values of a few times 
$10^6$ erg~cm$^{-2}$~s$^{-1}$ at chromospheric height levels.}
   { The main conclusions of our work are twofold:
firstly, for transverse, impulsive excitation, flux tubes/sheets with strong fields 
are more efficient than those with weak fields in providing acoustic flux
to the chromosphere. However, there is insufficient energy in the acoustic 
flux to balance the chromospheric radiative losses in the network, even
for the strong field case. Secondly, the acoustic emission from the interface 
between the flux concentration and the ambient medium decreases with the width 
of the boundary layer.}
  \keywords{Sun: magnetic field -- Sun: photosphere -- Sun: faculae, plages --
            Magnetohydrodynamic (MHD) -- Waves}
  \maketitle


\section{Introduction}
\label{sect_intro}

Quantitative studies of wave propagation in magnetically structured and 
gravitationally stratified atmospheres help to identify various physical 
mechanisms that contribute to the dynamics of the magnetic network on 
the Sun, and to develop diagnostic tools for the helioseismic exploration 
of such atmospheres. Magnetic fields play an important role in the generation 
and propagation of waves. The aim of this work is to attempt a better 
understanding of this process in the magnetized solar atmosphere. We have 
carried out a number of numerical simulations of wave propagation in a 
two-dimensional gravitationally stratified atmosphere consisting of individual 
magnetic flux concentrations representative of solar magnetic network 
elements of different field strengths.

While the magnetic field in the internetwork regions of the quiet Sun is mainly 
shaped by the convective-granular flow with a predominance of horizontal
fields and rare occurrence of flux concentrations surpassing 1~kG, the 
magnetic network shows plenty of flux concentrations at or surpassing this limit 
with a typical horizontal size-scale in the low photosphere of 100~km. These 
``magnetic elements'' appear as bright points in G-band images near disk center
and they can be well modeled as magnetic flux tubes and flux sheets. Their
magnetic field is mainly vertically directed and they are in a highly dynamical 
state 
\citep{muller1985,muller1994,berger1996,berger2001}.

Different from the shock induced \ion{Ca}{ii} $H_{2v}$ and $K_{2v}$ bright 
points in the cell interior, the network in the chromosphere is seen to be 
continuously bright 
\citep{lites1993,sheminova2005},
which asks for a steady heating mechanism. It is also seen that the \ion{Ca}{ii} H
and K line profiles from the network are nearly symmetric
\citep{grossmann1974}.

Several numerical investigations have been carried out to explain these
observations. Early works modelled the network as thin flux tubes and studied 
the transverse and longitudinal waves, which can be supported by them, excited 
by the impact of granules. These works failed to explain the persistent
emission that was seen in observations of the \ion{Ca}{ii} H and K lines. When
high frequency waves, generated by turbulence in the medium surrounding
flux tubes, were added 
\citep{hasan2000}, 
the observational 
signature of the modelled process became less intermittent and was in better 
agreement with the more steady observed intensity from the magnetic network. 
Later works examined mode coupling between transverse and longitudinal 
modes in the magnetic network, using the nonlinear equations for a thin flux
tube 
\citep[eg.][]{hasan2004}. 
All these studies modelled the network as consisting of thin-flux tubes, an 
approximation that becomes invalid at about
the height of formation of the emission peaks in the cores of the H and K
lines. Also, this approximation does not treat the dispersion
of magnetic waves caused by the variation of the magnetic field strength
across the flux concentration and it does not take into account the emission of
acoustic waves into the ambient medium.

Numerical simulations by 
\citet{rosenthal2002} and \citet{bogdan2003}, studied wave propagation in 
two-dimensional stratified atmospheres in the presence of a magnetic field. 
They recognized and highlighted the role of refraction of fast magnetic 
waves and the role of the surface of equal Alfv\'en and sound speed as a 
wave conversion zone, which they termed the magnetic canopy.
While the thick flux sheets of \citet{rosenthal2002} and \citet{bogdan2003} 
were a more realistic model for the network, they also assumed that the magnetic 
field was potential. Considering that the gas pressure, kinetic energy 
density, and the energy density of the magnetic field are all of similar magnitude 
in the photosphere, this assumption is probably not satisfied.

\citet{cranmer2005} modelled
the network as consisting of a collection of smaller flux tubes that are
spatially separated from one another in the photosphere. 
\citet{hasan2005}
performed MHD simulations of wave generation and propagation in an
individual magnetic flux sheet of such a collection and confirmed the
existence of magneto-acoustic waves in flux sheets as a result of the
interaction of these magnetic flux concentrations with the surrounding
plasma. They used a non-potential field to model the network. They
speculated that a well defined interface between the flux sheet and the
ambient medium may act as an efficient source of acoustic waves to the 
surrounding plasma. In a later paper, \citet{hasan2008} showed that 
the short period waves that are produced as a result of turbulent motions 
can be responsible for the heating of the network elements. 

\citet{cally2005,cally2007}
provided magneto-acoustic-gravity dispersion relations for waves in a
stratified atmosphere with a homogeneous, inclined magnetic field and
discussed the process of mode transmission and mode conversion. 
\citet{khomenko2008} 
presented results of nonlinear, two-dimensional, numerical
simulations of magneto-acoustic wave propagation in the photosphere and
chromosphere  in small-scale flux tubes with internal structure. Their
focus was on long period waves with periods of three to five minutes. 
\citet{steiner2007} 
considered magnetoacoustic wave propagation 
in a complex, magnetically structured, non-stationary atmosphere. They
showed that wave travel-times can be used to map the 
topography of the surface of thermal and magnetic equipartition ($\beta=1$) 
of such an atmosphere. 
\citet{hansteen2006} and 
\citet{pontieu2007} 
performed two-dimensional
simulations covering the solar atmosphere from the convection zone to the
lower corona. They showed how MHD waves generated by convective flows 
and oscillations in the photosphere turn into shocks higher up and produce 
spicules. 

Despite these efforts, the physical processes that contribute to the 
enhanced network emission are still not well understood. 
It is well known, that small scale magnetic elements have varying field strengths, 
ranging from hecto-gauss to kilo-gauss \citep{solanki1993,berger2004}. 
This suggests that the $\beta=1$ layer in these elements varies considerably 
in height, which in turn should affect the wave propagation in them
\citep{schaffenberger2005}. 
\citet{hasan2005} and \citet{hasan2008}, 
argue that the network is heated by the dissipation of magnetoacoustic 
waves. However, these works did not provide quantitative estimates of the 
energy flux carried by the waves. This is the main focus of the present investigation,
where we examine wave propagation in magnetic elements with different
magnetic field strengths.  We also study the effects of varying the
interface thickness between the flux sheet and the ambient medium on the
acoustic wave emission in the ambient medium. 

The outline of the paper is as follows. Section~\ref{sect_equilib}, discusses the
construction of the initial equilibrium model and Sect.~\ref{sect_meth} the boundary
conditions and method of solution for the simulation. In Sect.~\ref{sect_dyn}, the dynamics
and in Sect.~\ref{sect_energy}, the energetics is discussed. Section~\ref{sect_blayer}, 
discusses the effects of boundary layer thickness on the acoustic wave emission. 
Sect.~\ref{sect_summary} summarizes the results and Sect.~\ref{sect_conclusion}
contains the main conclusion and a discussion of the results.


\section{Initial equilibrium model}
\label{sect_equilib}

\begin{table*}
\caption{Equilibrium model parameters for the moderate and strong flux sheets.}
\label{parameters}
\centering
\renewcommand{\arraystretch}{1.2}
\begin{tabular}{llcccc}
\hline\hline
\multirow{2}{*}{Physical quantity} & & \multicolumn{2}{c}{Moderate field} 
                                                     & \multicolumn{2}{c}{Strong field} \\
\cline{3-6}
& & Sheet Axis & Ambient medium & Sheet Axis & Ambient medium\\
\cline{1-6}
\multirow{2}{*}{Temperature [K]} &  & 9142 & 9142 & 9142 & 9142 \\
 & & 4758 & 4758 & 4758 & 4758\\[0pt]
\cline{1-6}
\multirow{2}{*}{Density [g cm$^{-3}$]}     &  \rule{0pt}{10pt}
                                                                & $1.2 \times 10^{-12}$ &  $2.4 \times 10^{-12}$ 
                                                                & $1.2 \times 10^{-12}$ &  $6.0 \times 10^{-12}$ \\[0pt]
 &  & $1.4 \times 10^{-7}$ & $2.7 \times 10^{-7}$ & $1.4 \times 10^{-7}$ 
                                                                              & $6.7 \times 10^{-7}$\\
\cline{1-6}
\multirow{2}{*}{Pressure [dyn cm$^{-2}$]} &  & 1.3 & 2.7 & 1.3 & 6.6 \\
 &  & $4.2 \times 10^{4}$& $8.3 \times 10^{4}$ & $ 4.2 \times 10^{4}$ 
                                                                           & $20.6 \times10^{4}$ \\
\cline{1-6}
\multirow{2}{*}{Sound speed [km s$^{-1}$]} &  & 13.5 & 13.5 & 13.5 &  13.5\\
 &  & 7.1 & 7.1 & 7.1 & 7.1\\
\cline{1-6} 
\multirow{2}{*}{Alfv\'en speed [km s$^{-1}$]} & & 304 & 212 & 582 & 259 \\
 & & 6 & 0.9 & 12  & 0.6 \\
\cline{1-6}
\multirow{2}{*}{Magnetic field [G]} & & 119 & 117 & 229 & 225 \\
 & & 801 & 17 & 1601 & 18\\
\cline{1-6}
\multirow{2}{*}{Plasma $\beta$ [--]} & \rule{0pt}{10pt}
                                                      & $2.0 \times 10^{-3}$  & $5.0 \times 10^{-3}$ 
                                                      & $6.0 \times 10^{-4}$  & $3.0 \times 10^{-3}$\\
 &  & 1.7 & $7.4 \times 10^{3}$ & 0.4 & $1.6 \times 10^{4}$\\
\cline{1-6}
\end{tabular}
\end{table*}

The initial atmosphere containing the flux sheet is computed in cartesian 
coordinates using the numerical methods described in 
\citet{steiner1986} \citep[see also][]{steiner2007}.
The method consists of initially specifying a magnetic field configuration
and the pressure distribution in the physical domain. The magnetic field
can be written in terms of the flux function $\psi(x, z)$ as 
\begin{equation}
 B_x = -\frac{\partial \psi}{\partial z}, B_z = \frac{\partial \psi}{\partial x}\,.
\label{eq:mag_component}
\end{equation} 
The gas pressure as a function of height and field line (flux value),
$p(\psi,z)$, is given by,
\begin{eqnarray}
  p(\psi,z) = \left\{ \begin{array}{l l} 
           \displaystyle \frac{p_{\rm p}+p_{\rm c}}{p_{0}}(p_{0} 
                               + p_{2}\psi^{2})                                                   
         &\mbox{if}\quad  0 \le  \psi \le \psi_{1}, \\[3ex]
            \multicolumn{2}{l}{\displaystyle  \frac{p_{\rm p}+p_{\rm c}}
            {p_{0}} (a(\psi - \psi_{1})^{n} + b(\psi - \psi_{1})^2 }\\[2.0ex]
           + c(\psi - \psi_{1}) + d)                                                            
         &\mbox{if}\quad \psi_{1} < \psi < \psi_{2}, \\[2ex]
            \displaystyle \frac{p_{\rm p}+p_{\rm c}}{p_{0}} (p_{0} 
           + \frac{B_{0}^{2}}{8\pi})                                                           
         &\mbox{if}\quad \psi_2 \le \psi \le \psi_{\rm max}, 
\end{array} \right.
\label{eq:pressure_height}
\end{eqnarray}
where the constants a, b, c, and d are chosen such that the pressure and
its first derivative with respect to $\psi$ is a continuous function of
$\psi$ and where we choose $n=8$. The three cases in 
Eq.~(\ref{eq:pressure_height}) refer to the interior of the flux sheet,
its boundary layer, and the ambient medium from top to bottom, respectively.
The quantities $p_{\rm p}$, $p_{\rm c}$ and $p_{0}$ are defined as,
\begin{eqnarray}
  p_{\rm p} = p_{0,\rm p} e^{-z/H_{\rm p}},\\
  p_{\rm c} = p_{0,\rm c} e^{-z/H_{\rm c}},\\
  p_{0} = p_{0,\rm p} + p_{0,\rm c},
\label{eq:pressure_scale}
\end{eqnarray} 
where $H_{\rm p}$ and $H_{\rm c}$ are the photospheric and chromospheric
pressure scale heights, respectively. We choose 
$H_{\rm p} = 110$\,km and $H_{\rm c} = 220$\,km. $B_0$ is the magnetic field
strength on the axis of the flux sheet at the reference height $z=0$.  
The equation of motion along each field line reduces to the hydrostatic
equation
\begin{equation}
 \vec{B} \cdot [\nabla p - \rho \vec{g}] = 0,
\label{eq:eqn_motion}
\end{equation} 
which defines us the density distribution and with it the temperature field.
From the equation of motion perpendicular to $\vec{B}$ we obtain the electric
current density
\begin{equation}
 \vec{J} = \frac{1}{B^2} \vec{B} \times [\nabla p - \rho \vec{g}], 
\label{eq:electric_current}
\end{equation} 
which after some manipulation reduces to
\begin{equation}
  \displaystyle
 j_{y} = \left. \frac{\partial p}{\partial \Psi}\right|_z\,.
  \label{eq:eqn_volume_current}
\end{equation}
The new magnetic field configuration can be calculated from the current density 
using the Grad-Shafranov equation,
\begin{equation}
 \frac{\partial^{2} \psi}{\partial x^{2}} + \frac{\partial^{2} \psi}{\partial z^{2}} = 4 \pi j_{y}.
\label{eq:grad_shafranov}
\end{equation}

\begin{figure}
  \resizebox{\hsize}{!}{\includegraphics{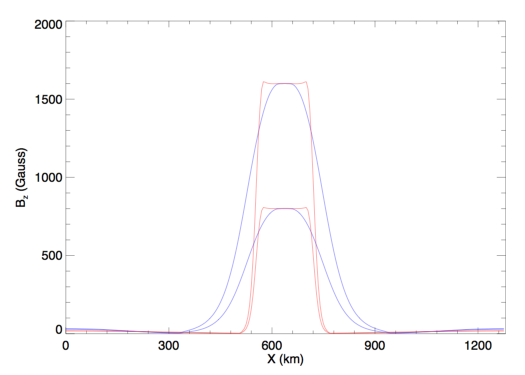}}
  \caption{Vertical component of the magnetic field at the base of the flux sheet,
  z=0. Red and blue curves correspond to field configurations with a sharp and a wide 
  interface to the weak-field surroundings, respectively. Each configuration is subdivided
  into a case of moderate field-strength with $\max(B_z)=800$~G and a case of strong
  field with $\max(B_z)=1600$~G.}
  \label{mag_vert_component}
\end{figure}

The above elliptic partial differential equation can be solved using
standard numerical methods with appropriate boundary conditions. In
practice we solve Eq.~(\ref{eq:grad_shafranov}) on a computational domain 
that consists of only half of the flux sheet of horizontal and vertical 
extensions of 640~km and 1600~km, respectively. 
This domain is discretized on a equidistant
rectangular mesh with a spacing of 5~km. The left side of the domain
corresponds to the axis of the flux sheet.  The value of $\psi$ is
prescribed at the left and the right side boundaries. At the top and bottom
boundaries we use the Neumann condition $\partial\psi/\partial z = 0$,
assuming that the horizontal field component vanishes at these two
boundaries. We treat  two different cases corresponding to field strengths
(at  $z = 0$) of 800~G and 1600~G, on the axis of the sheet. For each of
these cases we consider two boundary widths: 20~km and 80~km at the
reference height $z=0$. These widths can be obtained by choosing
appropriate values of $\psi_1$ and $\psi_2$ in Eq.~(\ref{eq:pressure_height}).

The vertical component of the magnetic field at $z=0$ is shown in 
Fig.~\ref{mag_vert_component} for the strong and moderate field cases. 
For the sharp interface (red curves) the vertical component of the magnetic 
field component drops sharply, whereas in the case of a wide interface 
(blue curves) the field decreases smoothly. 

The characteristic properties of the two models are summarized in 
Table~\ref{parameters}. The numbers in the first row of each quantity 
corresponds to the top
boundary ($z = 1600$~km) and the numbers in the second row corresponds to
the bottom boundary ($z = 0$~km). The temperature increases monotonically
from 4758~K in the photosphere to 9142 K in the chromosphere corresponding
to the sound speed variation from 7.1~km\,s$^{-1}$ to 13.5~km\,s$^{-1}$. The
density and pressure at the axis of the sheet is the same for both the
cases. We should mention that the ambient magnetic field is weak (of the order
of few tens of Gauss). As we go higher up in the atmosphere the flux sheet
expands and becomes uniform near the top with a average field strength of
118~G and 227~G for the moderate and strong field cases, respectively. The
plasma-$\beta$ on the sheet axis is 1.69 and 0.42 at the base for the moderate
and strong field cases.


\section{Method and boundary conditions}
\label{sect_meth}

Waves are  excited in the equilibrium magnetic field configuration through
a transverse motion of the lower boundary (similar to 
Hasan et al.  2005).
The system of MHD equations, given in conservation-law form for an inviscid adiabatic
fluid, is solved according to the method described in 
\citet{steiner1994}. 
These are the continuity,
momentum, entropy, and the magnetic induction equations. The unknown
variables are the density, $\rho$, the momenta, $\rho V_{x}$ and 
$\rho V_{z}$, where $V_x$ and $V_z$ are the horizontal and vertical components of
the velocity, the entropy per unit mass, $s$, and the magnetic
field, $B_{x}$ and $B_{z}$. The equation of state is that for the solar mixture with a 
constant mean molecular weight of 1.297. For the numerical integration, the system of MHD 
equations are transformed into a system of discrete finite volume equations. 
The numerical fluxes are computed based on the flux-corrected transport (FCT) scheme of 
\citet{oran+boris1987}. 
For the induction equation we use a constrained transport scheme 
\citep{devore1991}, 
which automatically keeps $\nabla \cdot \vec{B} = 0$. The time
integration is explicit. The scheme is of second order accuracy.

Transmitting conditions apply to the side boundaries set by constant extrapolation of the 
variables from the physical domain to the boundary cells. Constant extrapolation also 
applies to the horizontal component of the momentum at the top and bottom boundary and 
to the vertical component at the bottom boundary. The density in the top boundary cells is 
determined using linear log extrapolation, while at the bottom boundary hydrostatic extrapolation 
applies. For the temperature constant extrapolation is used. The horizontal component of the 
magnetic field at the top and bottom boundaries are set equal to the corresponding values at 
the preceding interior point. The vertical component of the magnetic field is determined by the 
condition $\nabla \cdot \vec{B} = 0$. 

The transverse velocity $V_{x}$ at $z = 0$ is specified as follows:
\begin{eqnarray}
  V_{x}(x,0,t) = \left\{ \begin{array}{l l} 
                 \displaystyle V_{0} \sin (2 \pi t / P) 
                 &\mbox{for}\quad 0 \le t \le P/2\,, \\[1ex]
                 \displaystyle 0
                 &\mbox{for}\quad 0 > t > P/2\,,
\end{array} \right.
\label{eq:velocity}
\end{eqnarray}
where $V_{0}$ denotes the amplitude of the horizontal motion and $P$ is the
wave period. This form was chosen to simulate the effect of transverse
motion of the flux sheet at the lower boundary.
For simplicity we assume that all points of
the lower boundary have this motion: this does not generate any waves
in the ambient medium, other than at the interface with the flux sheet. 
As a standard case in our simulation we use $V_{0} = 750$~m\,s$^{-1}$ 
and $P$ = 24~s following \citet{hasan2005}. This short period is motivated 
by the result of \citep{hasan2000} that high frequency waves would model 
the observational signature of wave heating less intermittent and thus in 
better agreement with the steady observed intensity from the magnetic network.
We consider a uniform horizontal displacement of the bottom boundary for
\emph{half a period} after which the motion is stopped 
\citep[this corresponds to the \emph{impulsive} case treated by][]{hasan2005}.
Such short duration motions are expected to be generated by the turbulent motion
in the convectively unstable subsurface layers where the flux sheet is rooted.
In terms of the analysis by \cite{cranmer2005} of the kinematics of
$G$-band bright points, this motion rather corresponds to a short, single step of 
their ``random walk phase'', for which these authors use a rms velocity of 
0.89~km\,s$^{-1}$ with a correlation time of bright-point motions of 60~s in 
accordance with the measurements of \cite{nisenson+al03}. The cases with 
higher velocities (see Table~\ref{tab:fluxes}) would rather be representative 
of the ``jump phase'' for which \cite{cranmer2005} use a velocity of 
5~kms$^{-1}$ with a duration of 20~s. This motion generates
magnetoacoustic waves in the flux sheet. We first examine
wave propagation and energy transport in a flux sheet with a sharp interface
for the moderate and the strong field cases. In Sect.~\ref{sect_blayer} we 
analyze the effect of varying the interface thickness.


\section{Dynamics}
\label{sect_dyn}

\subsection{\textit{Moderate field}} 
\label{sect_moderate}

\begin{figure}
  \centering
  \includegraphics[scale=.72]{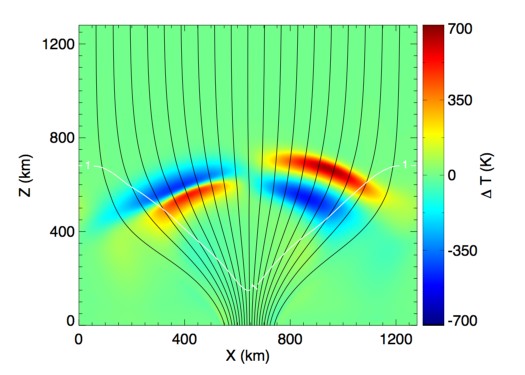}\\
  \includegraphics[scale=.72]{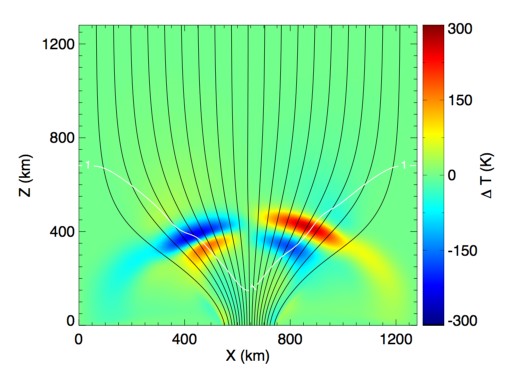}\\
  \includegraphics[scale=.72]{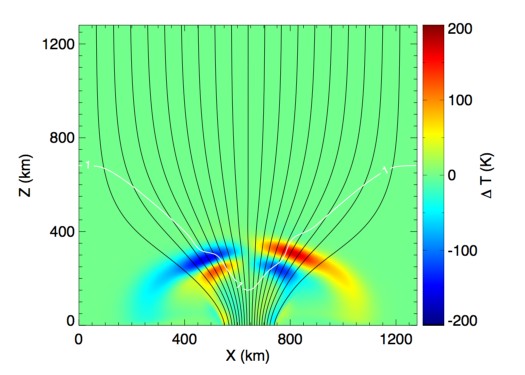}\\
  \includegraphics[scale=.72]{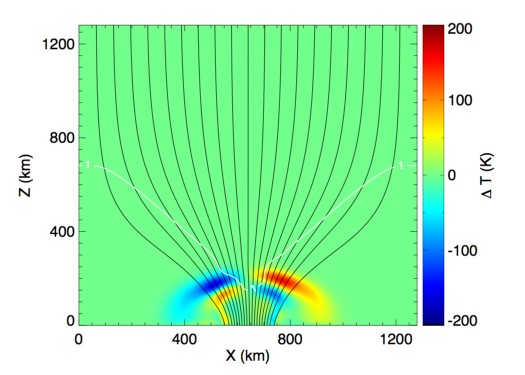}\\
  \caption{Temperature perturbations for the case in which the field strength
  at the axis at $z=0$ is 800~G (moderate field). The colors show the temperature
  perturbations at 40, 60, 80, and 120~s (from bottom to top) after initiation of an impulsive
  horizontal motion at the $z=0$ boundary of a duration of 12~s with an amplitude of 
  750~m\,s$^{-1}$ and a period of $P=24$~s. The thin black curves are field lines 
  and the white curve represents the contour of $\beta=1$.}
  \label{delta_t_moderate}
\end{figure}

\begin{figure*}
\centerline{
\begin{tabular}{cc}
  \includegraphics[scale=.90]{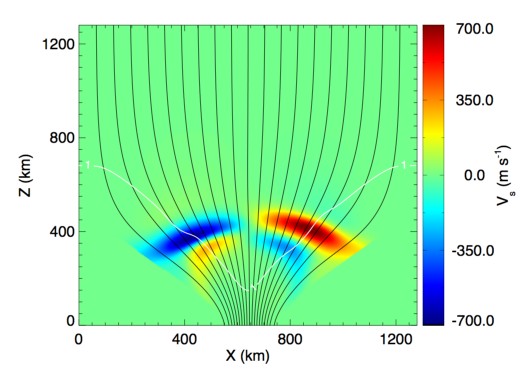} 
        & \includegraphics[scale=.90]{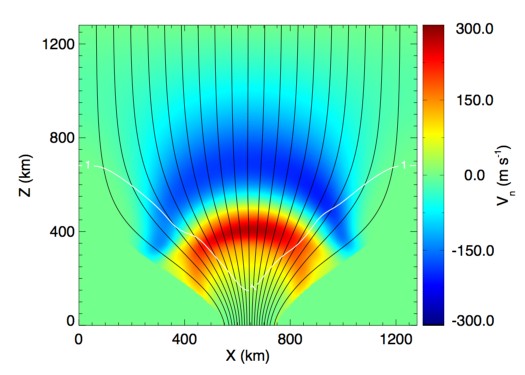} \\
  \includegraphics[scale=.90]{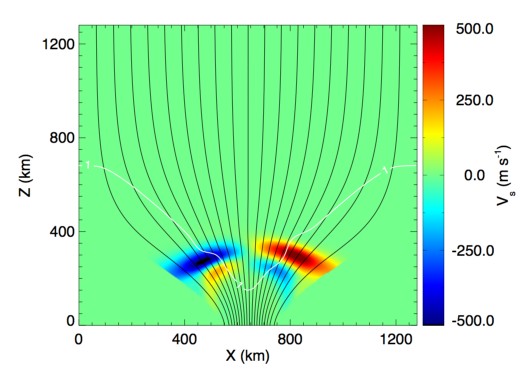} 
        & \includegraphics[scale=.90]{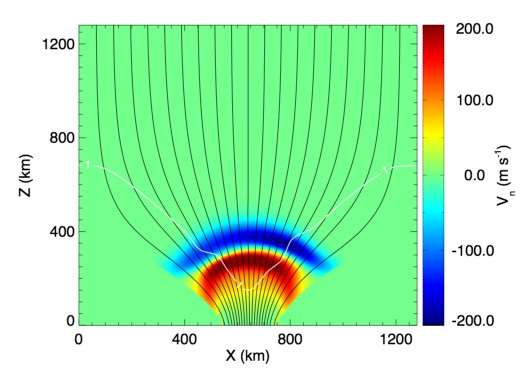} \\
  \includegraphics[scale=.90]{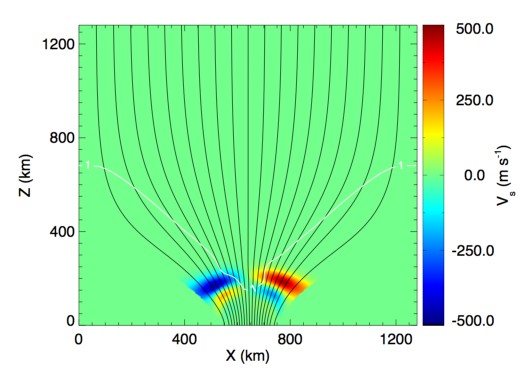} 
        & \includegraphics[scale=.90]{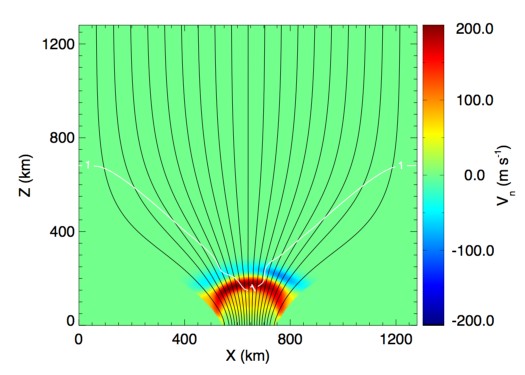} \\
  \mbox{a) $V_{s}$} & \mbox{b) $V_{n}$} 
  \end{tabular}
}
  \caption{Velocity components for the case in which the field strength at the
  axis at z=0 is 800~G (moderate field). The colors show the velocity components
  (a) $V_s$, along the field, and (b) $V_n$, normal to the field, at 40, 60, and 80~s 
  (from bottom to top)
  after initiation of an impulsive horizontal motion at the $z=0$ boundary of a duration 
  of 12~s with an amplitude of 750~m\,s$^{-1}$ and a period of $P=24$~s.  The thin 
  black curves are field lines and the white curve represents the contour of 
  $\beta=1$. The field aligned and normal components of velocity are not shown in 
  the regions where $B < 50$~G.}
  \label{velocity_moderate}
\end{figure*}

Let us consider a magnetic configuration in which the field strength at the
axis of the flux sheet at $z=0$ is 800~G. In this case the $\beta=1$ contour
is well above the bottom boundary in the atmosphere and hence all the 
magnetic field lines
emerging from the base of the sheet cross this layer at some height.  Waves
are excited at $z=0$, where $\beta >1$ (on the axis $\beta = 1.8$), in the
form of a fast (predominantly acoustic) wave and a slow 
(predominantly magnetic)\footnote{For brevity we call modes in the following 
simply acoustic and magnetic depending on the predominance of the thermal and 
magnetic nature of their restoring forces.}
wave, which propagate respectively at the sound and the Alfv\'en speeds. 
On the sheet axis, the acoustic and Alfv\'en speeds at $z=0$ are 7.1 and 
6.0~km\,s$^{-1}$, respectively (see Table~\ref{parameters}).  The fast wave 
is created due to compression and rarefaction of the gas at the leading and
trailing edge of the flux sheet, respectively: this can be clearly discerned 
in the snapshots of the temperature perturbation, $\Delta T$ (the temperature 
difference with respect to the initial value),  shown in  
Fig.~\ref{delta_t_moderate} at 40, 60, 80 and 120~s after start of the 
perturbation.  (These panels and panels in the following figures do not
show the full height range of the computational domain but up to 1280~km
only.) The black curves denote the magnetic field lines
and the white curve depicts the $\beta=1$ contour.  The perturbations are
$180^\circ$ out of phase on opposite sides of the sheet axis.  As these fast
waves travel upwards they eventually cross the layer of $\beta=1$, 
where they change their label from ``fast'' to ``slow'', without changing
their acoustic nature: this corresponds to  a ``mode transmission'' 
in the sense of \cite{cally2007}. 
The transmission coefficient depends (among others) on the ``attack angle" i.e.,
the angle between the wave vector and the local direction of the magnetic
field 
\citep{cally2007}. 
On the $\beta=1$ layer, away from the sheet axis,
where the wave vector is not exactly parallel to the magnetic field, we do
not have complete transmission of the fast wave to a slow wave.  Rather,
there is a partial conversion  of the mode from fast acoustic to
fast magnetic, so that the energy in the acoustic mode is
reduced correspondingly. 

\begin{figure}
  \centering
  \includegraphics[scale=.74]{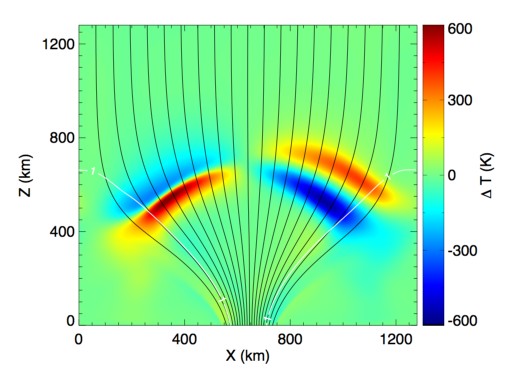}\\
  \includegraphics[scale=.74]{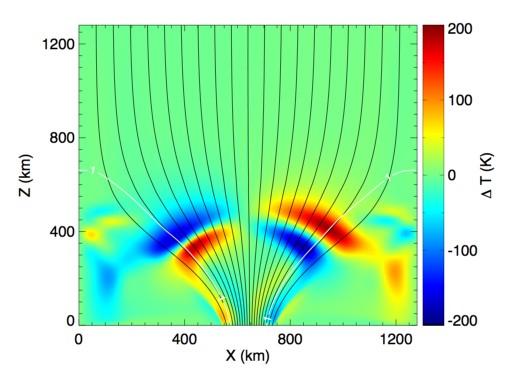}\\
  \includegraphics[scale=.74]{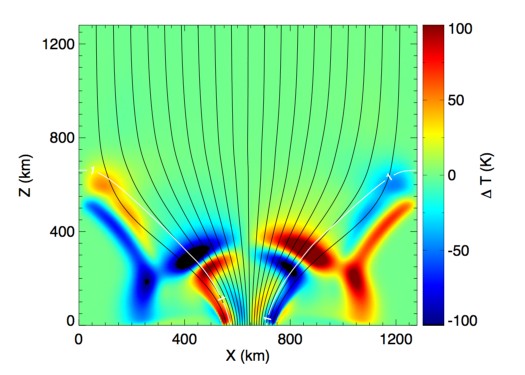}\\
  \includegraphics[scale=.74]{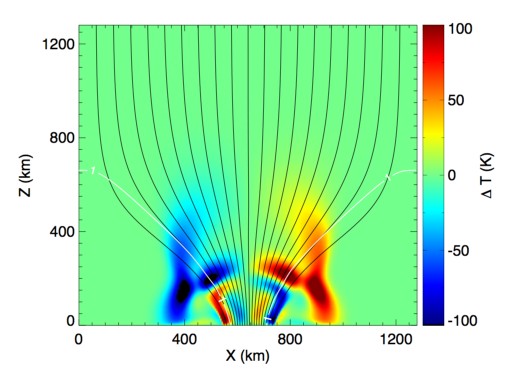}\\
  \caption{Temperature Perturbations for the case in which the field strength at
  the axis at z=0 is 1600~G (strong field) for times $t= 40$, 60, 80, and 120~s. 
  The coding corresponds to that of Fig.~\ref{delta_t_moderate}.}
  \label{delta_t_strong}
\end{figure}

Figures~\ref{velocity_moderate}a and \ref{velocity_moderate}b shows the 
velocity components in the flow parallel
($V_s$) and perpendicular ($V_n$) to the field, respectively. The velocity
components are shown only in regions where the field is greater than 50~G
since in the ambient medium with weak field this decomposition is no longer
meaningful. In general the waves possess both longitudinal and transverse
velocity components, but in regions where $\beta < 1$, the parallel
component essentially corresponds to the slow (acoustic) wave that is
guided upward along the field.  This correspondence can be seen by
comparing the parallel flow pattern (in Fig.~\ref{velocity_moderate}a) with the 
temperature perturbation in Fig.~\ref{delta_t_moderate}.  

The excitation at the bottom boundary also generates a slow (magnetic) wave
with velocity perturbations normal to the magnetic field line. In
order to visualize the slow wave, we show the velocity component normal to
the magnetic field in Fig.~\ref{velocity_moderate}b. The slow wave
also encounters the layer of $\beta=1$ and undergoes mode transmission and
conversion.  Above the layer of $\beta=1$, the transmitted wave is a fast mode,
which rapidly accelerates due to the sharp increase in Alfv\'en speed
with height.

\subsection{\textit{Strong field}}
\label{sect_strong}

\begin{figure*}
\centering
\begin{tabular}{cc}
  \includegraphics[scale=.90]{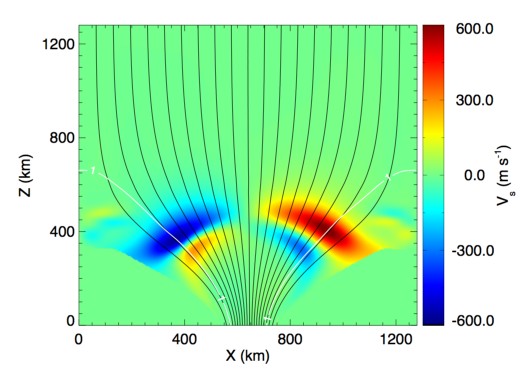} 
      & \includegraphics[scale=.90]{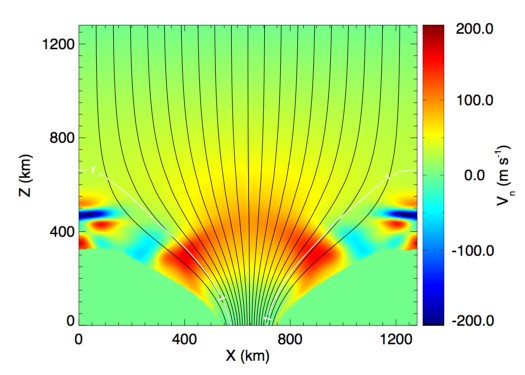} \\
  \includegraphics[scale=.90]{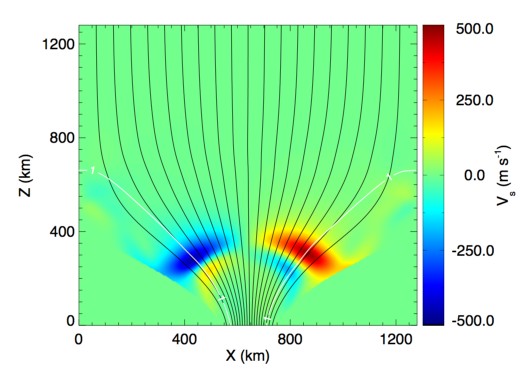} 
      & \includegraphics[scale=.90]{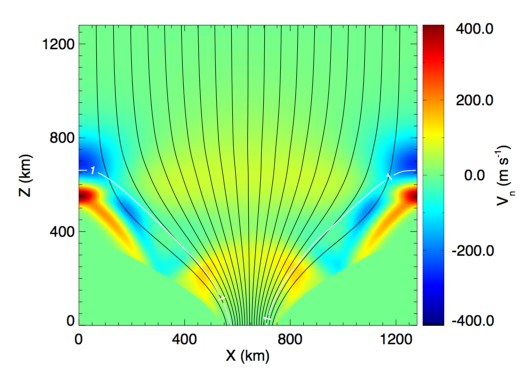} \\
  \includegraphics[scale=.90]{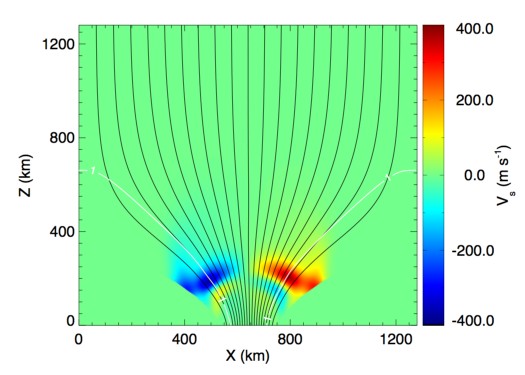} 
      & \includegraphics[scale=.90]{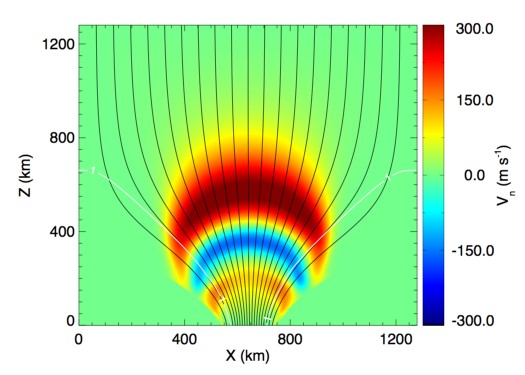} \\
  \mbox{a) $V_{s}$} & \mbox{b) $V_{n}$} 
\end{tabular}
  \caption{Velocity components for the case in which the field strength at the
  axis at z=0 is 1600~G (strong field) for times $t= 40$, 60, and 80~s. 
  The coding corresponds to that of Fig.~\ref{velocity_moderate}.}
  \label{velocity_strong}
\end{figure*}

We now consider the case in which the field strength on the sheet axis is
1600~G (at $z=0$). Here, the contour of $\beta=1$ approximately traces the boundary
of the flux sheet. The transverse motion of
the lower boundary generates slow (essentially acoustic) and fast (essentially 
magnetic) waves.  Since the contour of $\beta=1$ runs along the boundary of the
flux sheet, waves generated in the sheet that travel upwards do not encounter
this layer and hence do not undergo mode conversion.  Figure~\ref{delta_t_strong} 
shows the temperature perturbation $\Delta T$ at 40, 60, 80, 
and 120~s.\footnote{The temperature perturbations along the flux-sheet edges in the
wake of the slow acoustic wave (red and blue ridges along the left and right boundary
in the lower part of the flux sheet, respectively) do not pertain to a traveling
wave. They are due to the finite shift of the flux sheet with respect to the initial,
static configuration. This shift is compensated for by a corresponding shift of the
unperturbed solution for the computation of energy fluxes in Sects.~\ref{sect_energy} 
and \ref{sect_blayer}.} 
Figure~\ref{velocity_strong} shows the parallel and perpendicular components 
(with respect to the magnetic field) of the velocity.

The slow (acoustic) wave is guided upwards along the field  without
changing character. On the other hand, the fast wave, which can travel
across the field encounters the $\beta = 1$ contour at the boundary of the flux
sheet. As the fast wave crosses this layer, it enters a region of negligible field 
and hence gets converted into a fast (acoustic) wave. This can be easily seen in
the snapshot of temperature perturbations at an elapsed time of 40~s. The
fast wave in the low-$\beta$ region, which is essentially a magnetic wave,
undergoes mode conversion and becomes an acoustic wave, which creates
fluctuations in temperature visible as wing like features in the periphery of
the flux sheet between approximately $z=200$ to 500~km. The fast wave gets 
refracted  due to the gradients in Alfv\'en speed higher up in the atmosphere. 
Furthermore, similar to \citet{hasan2005}, we find that the interface between 
the magnetic flux sheet and the ambient medium is a remarkable source of 
acoustic emission. It is visible in Fig.~\ref{delta_t_strong} as a wave of
shell-like shape in the ambient medium that emanates from the base of the 
flux sheet and subsequently propagates, as a fast acoustic wave, laterally away 
from it.

Incidentally, the phase of transverse movement changes by $180^\circ$ between
the moderate and strong field case as can be seen comparing 
Fig.~\ref{velocity_moderate}b with
Fig.~\ref{velocity_strong}b. This is due to the development of
a vortical flow from the high pressure leading edge of the flux sheet to
the low pressure trailing edge that develops in the high-$\beta$ photospheric
layers of the moderately strong flux sheet but is largely suppressed in the
strong field case, where it is from the beginning preceded by the fast
(magnetic) wave that emerges right from the initial pulse. The development of
a vortical flow in the moderate field case was also noticed in \citet{hasan2005}.

Besides the fast and slow acoustic and the fast magnetic wave that emanate directly
from the initial perturbation there is also a slow magnetic mode from this
source, which propagates in the high-$\beta$ surface layer of the flux sheet.
It is visible in Fig.~\ref{velocity_strong}b as the yellow/red crescent-shaped
perturbation, which trails the red and blue crescents pertaining to the fast
magnetic mode. Different from the latter, which is maximal at the flux-sheet 
axis, the slow mode has maximal amplitude in the weak-field boundary-layer of
the flux concentration. A similar but acoustic slow surface mode was found
by \citet{khomenko2008} when the driver was located in the high-$\beta$ layers
of the flux concentration. Here, this mode generates a remarkable amount of acoustic
emission to the ambient medium as will be seen in Sect.~\ref{sect_blayer}.

In a three-dimensional environment there would in general appear a third,
intermediate wave type, additional to the slow and fast mode, depending on the
geometry of the magnetic field and the initial perturbation.  Correspondingly,
we may expect mode coupling between all three wave types. In the presence 
of gravitational stratification, the $\beta=1$ surface (more precisely 
the surface of equal Alfv\'en and sound speed) would still constitute the
critical layer for mode coupling so that we could expect scenarios not radically
different from but more complex than those of Secs.~\ref{sect_moderate} and 
\ref{sect_strong}.

 
\section{Energy transport}
\label{sect_energy}

We now consider the transport of energy in the various wave modes. Using
the full nonlinear expression for the energy flux, it is not easily
possible to identify the amount of energy carried by the magnetic and
acoustic waves. Following 
\citet{bogdan2003}, 
we instead consider the wave flux using the expression given by 
\citet{bray1974} 
that represents the net transport of
energy into the atmosphere:
\begin{equation} \vec{F}_{\rm wave} =
\Delta p\vec{V} + \frac{1}{4 \pi}(\vec{B}_{0} \cdot \Delta
\vec{B})\vec{V} - \frac{1}{4 \pi}(\vec{V} \cdot \Delta
\vec{B})\vec{B}_{0} .
\label{wave_energy}
\end{equation} 
The first term on the right hand side of the equality sign is the net acoustic 
flux, and the last two terms are the
net Poynting flux. The operator $\Delta$ gives the perturbations in the
variable with respect to the initial equilibrium solution and $B_0$ 
refers to the unperturbed magnetic field. 

\begin{figure*}
\centering
\begin{tabular}{cc}
  \includegraphics[scale=.90]{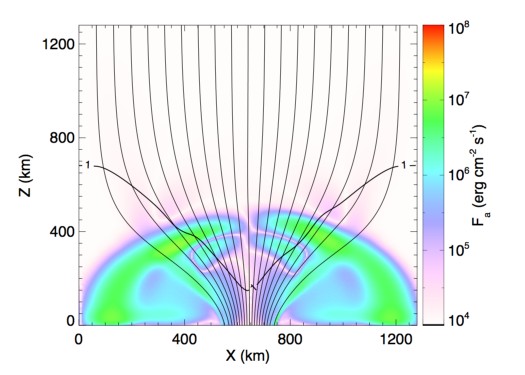} &
  \includegraphics[scale=.90]{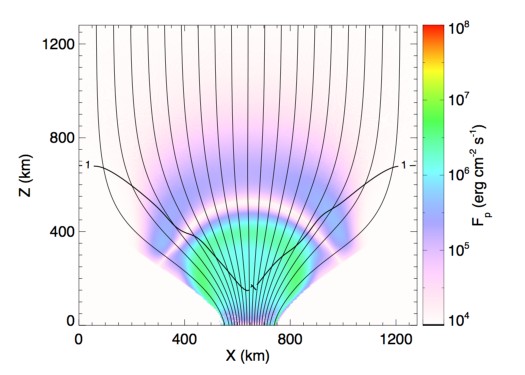}\\
  \includegraphics[scale=.90]{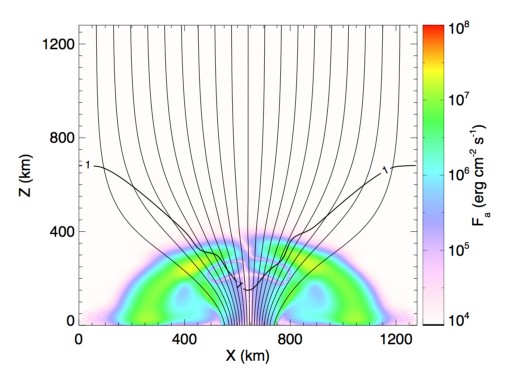} &
  \includegraphics[scale=.90]{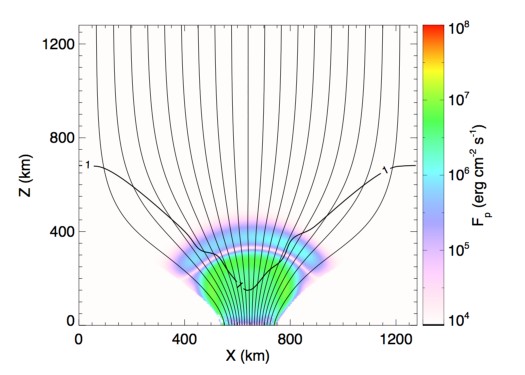} \\
  \includegraphics[scale=.90]{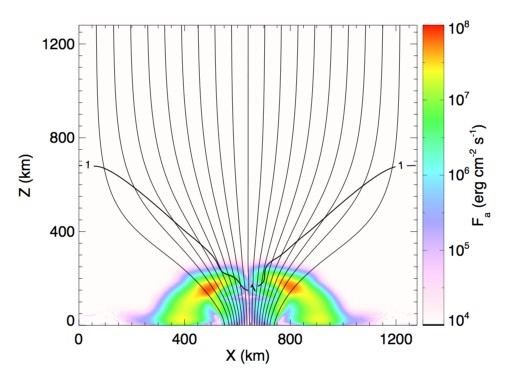} &
  \includegraphics[scale=.90]{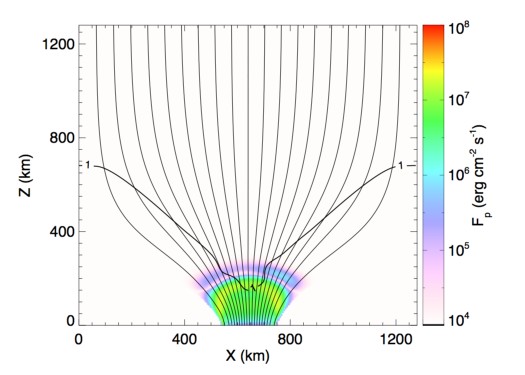} 
  \\
  \mbox{a) Acoustic} & \mbox{b) Poynting} 
\end{tabular}
  \caption{Wave-energy fluxes (absolute values) for the case in which the field strength 
  at the axis at $z=0$ is 800~G (moderate field). The colors show (a) the acoustic
  flux, and (b) the Poynting flux, at 40, 60, and 80~s (from bottom to top) after initiation of 
  an impulsive horizontal motion at the $z=0$ boundary of a duration of 12~s with an amplitude 
  of 750~m\,s$^{-1}$ and a period of $P=24$~s.  The thin black curves are field lines 
  and the thick black curve represents the contour of $\beta=1$. The Poynting fluxes are 
  not shown in the ambient medium where $B < 50$~G.}
  \label{fluxes_moderate}
\end{figure*}

Figures \ref{fluxes_moderate}a and \ref{fluxes_moderate}b show the magnitude 
of the acoustic (left panels) and the
Poynting flux (right panels) at 40, 60, and 80~s (from bottom to top) for the moderate
field case. Since in the ambient medium the field strength is weak, the Poynting 
fluxes are not shown in this region. The Poynting flux is essentially the wave energy 
that is carried by the magnetic mode, which as expected, is localized to the 
flux sheet. 
On the other hand, the energy transport in the acoustic-like component is more
isotropic. At $t=40$~s, we find from Fig.~\ref{fluxes_moderate}a that the wave 
has just crossed the $\beta=1$ contour. Thereafter, it propagates as a slow wave
guided along the field at the acoustic speed within the flux sheet and as a fast
spherical-like wave in the surrounding quasi field-free medium. Inside the flux sheet, 
the energy in the magnetic component (Poynting flux)  and the acoustic component is 
of the same order of magnitude.

\begin{figure*}
\centering
\begin{tabular}{cc}
  \includegraphics[scale=.90]{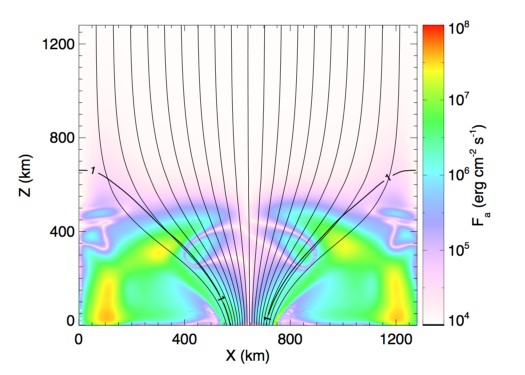} & 
  \includegraphics[scale=.90]{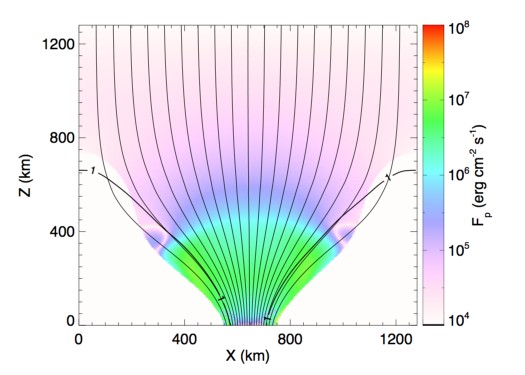} \\
  \includegraphics[scale=.90]{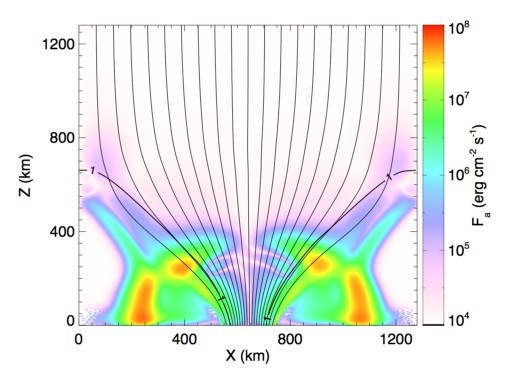} & 
  \includegraphics[scale=.90]{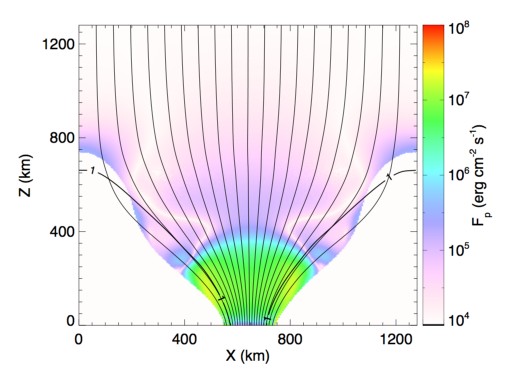} \\
  \includegraphics[scale=.90]{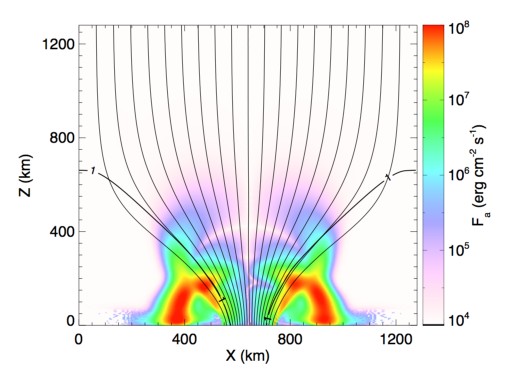} & 
  \includegraphics[scale=.90]{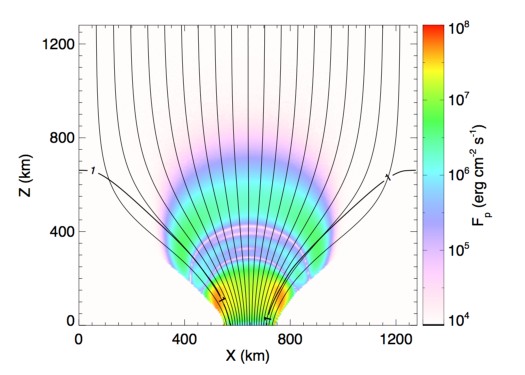} \\
  \mbox{a) Acoustic} & \mbox{b) Poynting} 
  \end{tabular}
  \caption{Wave-energy fluxes for the case in which the field strength at the
  axis at $z=0$ is 1600~G (strong field) for times $t= 40$, 60, and 80~s.
  The coding corresponds to that of Fig.~\ref{fluxes_moderate}. The Poynting 
  fluxes are not shown in the ambient medium where $B < 200$~G.}
  \label{fluxes_strong}
\end{figure*}

A comparison of these results with those for the strong field case 
(Fig.~\ref{fluxes_strong}) shows that in the latter case energy is transported 
by the fast wave much more rapidly, especially in the central regions of the flux sheet.
This is due to the sharp increase of the Alfv\'en speed with height above $z > 200$~km. 
At $t=40$~s we find that the wave front associated with the magnetic component
has already reached a height of about 500~km (close to the sheet axis), while
the acoustic wave reaches this level only at about $t=80$~s.  

From the contour plots of Figs.~\ref{fluxes_moderate} and \ref{fluxes_strong}, 
we see that the fluxes in the ambient medium for the strong field case 
is still close to  $10^{8}$~erg~cm$^{-2}$~s$^{-1}$, while for the moderate field, 
it is almost an order of magnitude less, suggesting that the flux sheets with strong 
fields are a more efficient source of acoustic fluxes into the ambient medium. 
The ``mode transmission'' from fast (acoustic) to slow (acoustic) that takes 
place in the case of a moderate field, as explained in Sect.~\ref{sect_dyn}, can be seen 
in Fig.\ref{fluxes_moderate}a. Since the ``attack angle'' in this case is close to zero, 
a significant amount of acoustic transmission takes place across the layer of $\beta=1$. 
Another feature that we see in 
the plots of wave-energy fluxes is the ``mode conversion'' that takes place in the strong field 
case. The fast magnetic wave, which is generated inside the flux sheet can travel across the 
magnetic field. This mushroom like shape, which is expanding, can be easily discerned in the 
40~s snapshot of the Poynting flux shown in Fig.~\ref{fluxes_strong}b. As this wave crosses 
the $\beta=1$ contour, it is converted into a fast (acoustic) wave. The wing like feature that 
can be seen in the 60~s snapshot of the acoustic fluxes (Fig.~\ref{fluxes_strong}a) are due to
the fast waves that have a moment ago undergone a ``mode conversion'' from magnetic to acoustic. 

\begin{figure*}
  \begin{center}
    \includegraphics[scale=.8]{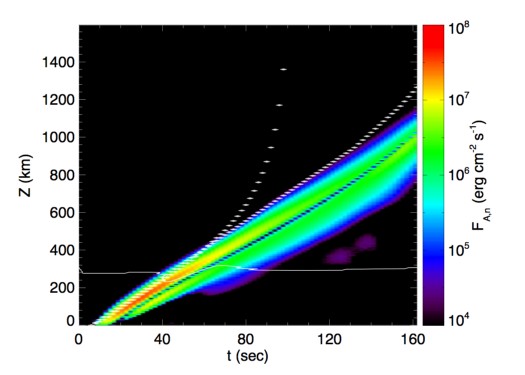}
    \includegraphics[scale=.8]{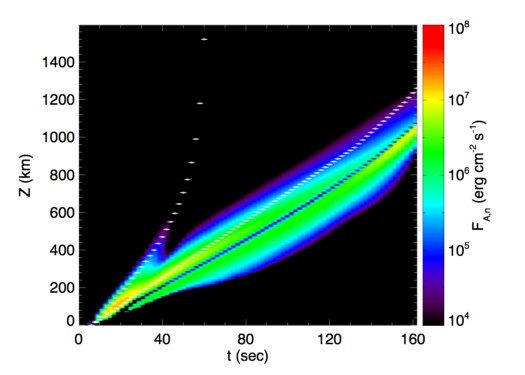}
  \end{center}
  \caption{The field aligned positive (upwardly directed) component of acoustic 
  wave-energy flux as a function 
  of time on a field line on the left side of the axis that encloses a fractional flux of
  50\%. The left panel represents the case in which the field strength at the
  axis at $z=0$ is 800~G (moderate field) and the right panel represents the strong 
  field case (1600~G).}
  \label{aco_para}
\end{figure*}

\begin{figure*}
  \begin{center}
    \includegraphics[scale=.8]{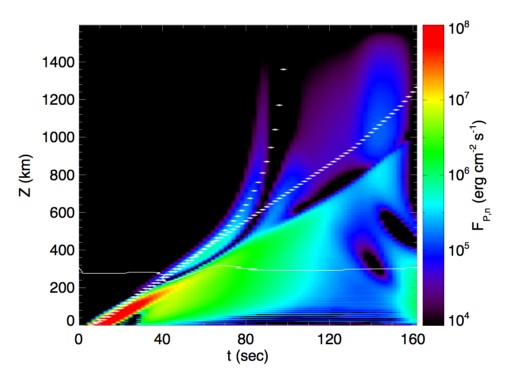}
    \includegraphics[scale=.8]{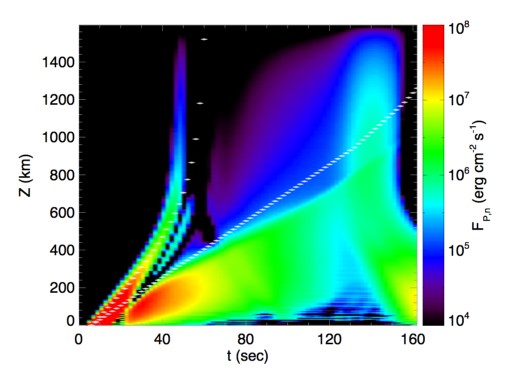}
  \end{center}
  \caption{The field aligned positive (upwardly directed) component of the Poynting flux 
  as a function 
  of time on a field line on the left side of the axis that encloses a fractional flux of
  50\%. The left panel represents the case in which the field strength at the
  axis at $z=0$ is 800~G (moderate field) and the right panel represents the strong 
  field case (1600~G).}
  \label{mag_para}
\end{figure*}

Next we consider a field line to the left of the flux sheet axis, which encloses a
fractional flux of 50\%. The field aligned and the normal component
of the wave-energy fluxes are calculated along this particular field line.
Fig.~\ref{aco_para} shows the positive, field aligned component of acoustic flux 
for the moderate and strong field case as a function of time and spatial coordinate
$z$ along the field line. The 
dotted curves in the figure show the space time position of a hypothetical wavefront 
that travels with Alfv\'en speed (steeper slope) and sound speed along this magnetic 
field line. With the help of this plots it is easy to separate the energy fluxes that is 
carried by the slow and the fast wave modes. The evolution of the $\beta=1$ layer 
is shown for the moderate field case. The perturbation of this layer as the wave 
crosses it can be seen clearly around 40 seconds. It moves down due to the 
decrease in pressure caused by the rarefaction front and then moves up when 
the compression front passes it. Most fraction of the flux lies parallel to the 
 line that corresponds to the hypothetical 
acoustic wave, which is a slow mode in the region where $\beta<1$. 
In the strong field case, above approximately 
$z=800$~km and for times $t \ga 120$~s,  the acoustic flux carried in the 
compressive (trailing) phase starts to catch up the slightly slower moving expansive 
phase and the flux gets confined into a narrow shock forming region. This is
also visible in the case of the moderately strong field. This behavior is not
present along the corresponding field line to the right of the sheet axis (not shown
here), where the compressive phase is leading so that the compressive
and expansive phase of the perturbation slightly diverge with time.

\begin{table*}
\centering
\caption{Temporal maximum of the horizontally averaged, vertical component of the 
wave-energy fluxes for the strong field case}.
\label{tab:fluxes}
\begin{tabular}{lrrrr|rrr}
\hline\hline
&& \multicolumn{3}{c}{$F_{A,z}$ ($10^{6}$~erg~cm$^{-2}$~s$^{-1}$) } 
& \multicolumn{3}{c}{$F_{P,z}$ ($10^{6}$~erg~cm$^{-2}$~s$^{-1}$)\rule[-4pt]{0pt}{14pt}}  \\ 
\cline{3-8}
\multicolumn{2}{l}{Initial Excitation\rule{0pt}{10pt} }& z=100~km & z=500~km & z=1000~km 
& z=100~km & z=500~km & z=1000~km  \\
\hline
0.75~km\,s$^{-1}$, &   24s \rule{0pt}{10pt}& 11.36  & 1.96  & 1.33 & 29.38 & 1.08 & 0.14\\
0.75~km\,s$^{-1}$, & 120s & 35.75 & 27.70 & 4.02 & 134.29 & 0.79 & 0.07\\
0.75~km\,s$^{-1}$, & 240s & 20.90 & 8.58 & 3.30 & 131.84 & 0.36 & 0.02\\
1.50~km\,s$^{-1}$, &   24s & 44.55  & 7.68  & 3.34 & 115.79 & 4.29 & 0.57\\
3.00~km\,s$^{-1}$, &   24s & 168.41 & 30.40 & 6.22 & 434.03 & 16.90 & 2.31\\
\hline
\end{tabular}
\end{table*}

The acoustic fluxes  are of the order $10^{7}$ ergs~cm$^{-2}$~s$^{-1}$. 
The Poynting fluxes carried by the fast mode in this region can be identified 
by the coloured contours that gather along the dotted curves corresponding 
to the hypothetical Alfv{\'e}n wave (Fig.~\ref{mag_para}). Comparing 
the two fluxes (Fig.~\ref{mag_para} with Fig.~\ref{aco_para}), 
it is clear that the acoustic flux carried by the slow mode is
larger than the Poynting flux, especially in the moderate field case.
The Poynting flux rapidly weakens with time and height because it is not 
guided along the field lines like the slow mode but rapidly diverges across
the field and part of the Poynting flux converts to acoustic again as explained
in Sect.~\ref{sect_dyn}. Also from Fig.~\ref{mag_para} it can be seen that
while the magnetically dominated fast mode starts right from the excitation level 
at $z=0$ in the strong field case, it starts in the weak field case 
only after about 40~s  when the fast acoustic wave reaches the conversion
layer where $\beta \approx 1$ and partially undergoes mode conversion. Therefore,
the fast (magnetic) mode is from the beginning weaker in the moderate as compared
to the strong field case.

Table~\ref{tab:fluxes} shows the temporal maximum of the horizontally averaged 
vertical components of acoustic and Poynting fluxes at three different heights for
the strong field case. We have considered three different amplitudes and periods
for the initial excitations. Although the field aligned acoustic fluxes on the 
specific field line considered in Fig.~\ref{aco_para} reach values of the order of
$10^{7}$~erg cm$^{-2}$~s$^{-1}$ at a height of z=1000~km,  the horizontally 
averaged fluxes are typically an order of magnitude less, depending upon the 
amplitude of the initial excitation.
The Poynting fluxes shown in the table are the maximum value that the fluxes 
reach in the interval between the start of the simulation until the time when the 
fast wave reaches the top boundary (around 60s). Hence these fluxes are due 
to the fast mode for $z=500$~km and $z=1000$~km, since within this time limit the 
slow mode has not yet reached these heights.
The Poynting fluxes associated with the fast mode are relatively lower in magnitude 
compared to the acoustic fluxes. It should be noted that there is also considerable Poynting 
flux associated with the slow mode, since these waves also perturb the magnetic field.
The acoustic fluxes of the moderate field case reach only less than 70\% of that of the
strong field configuration and the Poynting fluxes are negligible in in this case.

 
\section{Effects of the boundary-layer width}
\label{sect_blayer}

We now study the acoustic emission of the magnetic flux concentrations 
into the ambient medium by varying the width of the boundary  layer 
between the flux sheet and
ambient medium.  This is carried out by comparing the result of
simulations with a sharp interface of width 20~km to that with a
width of 80~km (see Fig.~\ref{mag_vert_component}), where the
width can be varied by choosing appropriate values of $\psi_1$ and
$\psi_2$ in Eq.~(\ref{eq:pressure_height}).

We examine the acoustic emission from the two peripheral (control) field 
lines to the left and to the right of the flux-sheet axis that encompass 
90\%  of the magnetic flux. These correspond to the outermost field 
lines that are plotted in Figs.~\ref{delta_t_moderate} to \ref{fluxes_strong}.
These field lines are located in the high-$\beta$ region 
with $\beta >> 1$, all the way from the base to the merging height,
where the flux sheet starts to fill the entire width of the computational
domain. The acoustic emission from the peripheral field
line to the right and to the left of the flux-sheet axis is practically 
identical.

\begin{figure*}
  \begin{center}
    \subfigure{\includegraphics[scale=0.8]{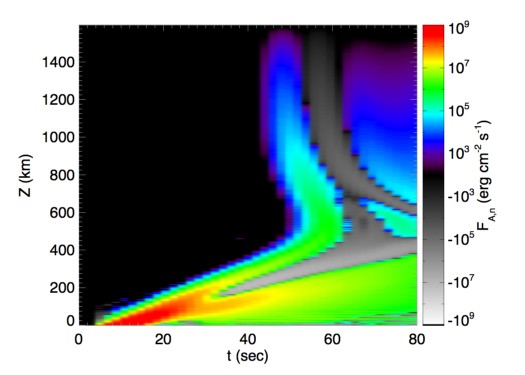}}
    \subfigure{\includegraphics[scale=0.8]{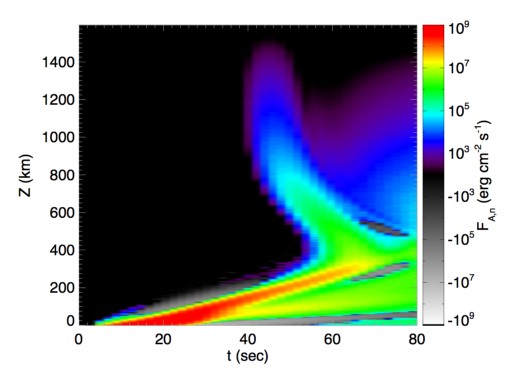}}
  \end{center}
  \caption{Acoustic flux perpendicular to the peripheral field lines that encompass 90\%
  of the magnetic flux as a function of time and height along the field line. Only the
  outwardly directed flux is shown. Left: strong field
  case with a sharp interface between flux-sheet interior and ambient medium. Right:
  strong field case with a wide interface.}
  \label{interface_1600}
\end{figure*}

Figure~\ref{interface_1600} shows the acoustic emission from the
flux sheet into the ambient medium for the peripheral field line to
the left of the
flux sheet with the strong field ($B_0=1600$~G) and the cases of the 
sharp interface (left panel) and the wide interface (right panel).
Concentrating on the case with the sharp interface first, we see that
acoustic flux is initially generated by the fast mode that stems from the 
transversal motion of the flux sheet to the right hand side, which causes a 
compression and expansion to the right and left side of the flux-sheet edge, 
respectively. This movement generates a net acoustic flux away from the flux 
sheet on both sides. It is visible in Figs.~\ref{delta_t_strong} and 
\ref{fluxes_strong} as the shell-like antisymmetric wave that emanates 
from the base of the flux sheet propagating into the ambient medium. 
At a height of $z=100$~km the peak value of this flux
is $3\times 10^8$~erg~cm$^{-2}$~s$^{-1}$ for the sharp interface but only
$1.2\times 10^8$~erg~cm$^{-2}$~s$^{-1}$ for the wide interface. This
is because the sharp interface acts like a hard wall that pushes against
the ambient medium, while the wide interface is more compressible and 
acts more softly.

\begin{table*}
\centering
\caption{Total acoustic emission from the flux sheet into the
ambient medium for different boundary layer widths.}
\label{tab:boundary_layer}
\begin{tabular}{lrr}
\hline\hline
Boundary-layer width & \multicolumn{2}{c}{Total acoustic emission\rule{0pt}{9pt}}\\
& \multicolumn{2}{c}{(10$^{16}$ erg\,cm$^{-1}$)}\\
\hline
20~km (sharp)  &\rule{30pt}{0pt} 23.40 &\rule{0pt}{9pt}\\
40~km (medium) & 13.22 &\\
80~km (wide)   &  8.13 &\\
\hline
\end{tabular}
\end{table*}

Near the flux-sheet boundary this wave seamlessly connects to the tips of 
the crescent-like fast (magnetic) mode of the flux-sheet interior as can 
be best seen when comparing the first two snapshots
of Figs.~\ref{delta_t_strong} and \ref{velocity_strong}b.
There, acoustic flux is generated by continuous
leakage and conversion from the magnetic mode, giving rise to the steeper
of the two horizontally running, inclined ridges of acoustic flux, visible
in the lower part of both panels of Fig.~\ref{interface_1600}. This leakage 
is more efficient
in the case of the wide interface than in the case to the sharp interface so 
that the corresponding ridge extends over a longer period of time in the
former compared to the latter case. However, it cannot compensate
for the larger initial flux that emanates from the more confined (sharp) 
boundary.

Starting at about $t=25$~s in case of the sharp interface, one can see a less 
steep and weaker branch of acoustic flux that is connected to the slow (magnetic) 
mode that propagates in the high-$\beta$ boundary-layer of the flux sheet. Obviously
it creates a non-negligible source of acoustic flux to the ambient medium.
It is also present in case of the wide interface.

The two horizontally running ridges of acoustic flux in the case of
the sharp interface (Fig.~\ref{interface_1600}, left) is slightly more
inclined compared to the case with the wide interface 
(Fig.~\ref{interface_1600}, left), where the peripheral (control) field line expands 
more in the horizontal direction so that the wave travels a longer distance 
to reach it.

At approximately $t=45$~s we start to see acoustic flux appearing at a height 
of about $z=1000$~km. This flux originates from the refracted fast (magnetic) 
wave within the flux sheet. Since this wave quickly accelerates and refracts 
with height, it reaches the flux-sheet boundary sooner at $z=1000$~km than in 
the height range $500$~km $< z < 800$~km. This wave undergoes conversion from 
fast, predominantly magnetic to fast, predominantly acoustic as it crosses
the region where $\beta = 1$. Because it travels essentially perpendicular
to the field near the flux-sheet boundary, the conversion is particularly 
efficient. While this
ridge of acoustic flux originates from the leading phase of the fast
(magnetic) wave that corresponds to a movement to the right 
(red big crescent in the 40~s snapshot of Fig.~\ref{velocity_strong}b),
a second, parallel running negative ridge, stems from the
following phase, corresponding to a movement to the left
(blue crescent in the 40~s snapshot of Fig.~\ref{velocity_strong}b).

Table~\ref{tab:boundary_layer} shows the total acoustic emission to the ambient
medium, still from and perpendicularly across the field lines that encompasses 
90\% of the total magnetic flux for cases with 3 different boundary layer widths.
The energy is computed by integration of the perpendicular flux along the peripheral
control field lines to the left and to the right over the full height range of the 
computational domain and from $t=0$~s to $t=64$~s for unit width.
The total acoustic energy leaving the flux sheet with the wide interface is only 
35\% of that with the sharp interface. In this sense, a flux sheet with a 
sharp interface is more efficient in providing acoustic flux to the ambient 
medium than a flux sheet with a wide interface as conjectured by
\citet{hasan2005}

 
\section{Summary}
\label{sect_summary}

This work is an extension of the previous work done by \citet{hasan2005}.
Wave excitation occurs in a magnetic flux concentration by a transverse motion 
of the base. The present work extends the previous calculations to the case of a 
flux sheet with moderate field strength.  In addition, a new feature of the present 
work is that we estimate the energy carried by the waves and we examine the 
effect of varying the thickness of the tube-ambient medium interface on the 
acoustic emission.

We have found that the nature of the modes
excited depends upon the value of $\beta$ in the region where the driving
motion occurs.  When $\beta$ is large, the slow wave is a transverse magnetic
mode that propagates along the field lines and undergoes mode transmission as
it crosses the $\beta=1$ layer. In this case, the wave only changes label from
slow to fast, but remains magnetic in character throughout the flux sheet. 
The fast mode, which propagates almost isotropically, undergoes both
mode conversion and transmission at the $\beta=1$ surface depending on the
``attack angle'', the angle between the wave vector and the magnetic field.  
On the other hand, in the case of a strong magnetic field (low-$\beta$ case), 
where the level of $\beta=1$ is below the driving region, the
fast (magnetic) and slow (acoustic) modes propagate through the flux sheet
atmosphere without changing character.

We find that the magnetically dominated  fast wave within the low-$\beta$ region
of the flux sheet undergoes strong refraction so that it finally leaves the flux sheet
in the lateral direction, where it gets partially and mainly converted to a fast, 
acoustically dominated wave. This effect is particularly visible in the case of a 
flux sheet with strong magnetic field.

We also see an asymmetry in the wave structure on both sides of the flux
sheet axis. This comes because the leading front of the predominantly acoustic
mode is compressional on the one hand side and expansive on the other side
and vice versa for the following phase. Since the compressive phase travels faster
as the sound speed is larger, the two phases move either apart from each other
or converge. In principle, this asymmetry should give rise to observable signatures.

Recent observations of the chromospheric network are suggestive of \ion{Ca}{ii}
network grains associated with plasma with  quasi-steady heating at heights
between 0.5 and 1~Mm inside magnetic flux concentrations
\citep{hasan2008}. 
Let us now estimate the acoustic energy flux transported into the
chromosphere through a single short duration pulse as has been treated in
the present paper (the magnetic modes are almost incompressible and not efficient
for heating the atmosphere). We consider the energy flux at a height of
1000~km.  For the strong and moderate field cases, the maximum values of 
the acoustic fluxes at  $z=1000$~km are
$\sim$15$\times$$10^{6}$~erg~cm$^{-2}$~s$^{-1}$ and 
$\sim$5$\times$$10^{6}$~erg~cm$^{-2}$~s$^{-1}$, respectively. 
However, it should be noted that although the fluxes can reach values up to 
$10^{7}$~erg~cm$^{-2}$~s$^{-1}$, the spatially averaged values are much 
less. From Table~\ref{tab:fluxes} we obtain for the strong field case 
a temporal maximum of the horizontally averaged acoustic flux at 
$z=1000$~km of a few times $10^{6}$~erg~cm$^{-2}$~s$^{-1}$, depending on 
the excitation amplitude and period. 

We see that the strong field configuration is a more efficient source of
acoustic waves in the ambient medium compared to the weak to
moderate field configurations. For the cases considered here, they differ
by almost a factor of two. The width of the transition
layer between the flux sheet and the ambient medium has significant effect
on the acoustic wave emission as was initially conjectured by 
\citet{hasan2005}. 
Our quantitative calculations substantiate this hypothesis: a flux sheet with
a sharp interface emits almost four times the energy emitted by a flux sheet 
with a wide interface.


\section{Discussion and conclusion}
\label{sect_conclusion}

The energy losses in the magnetic network at chromospheric heights are of the order of 
$10^{7}$~erg~cm$^{-2}$~s$^{-1}$.  Even though the acoustic energy flux produced by the
transverse excitation movement can temporarily reach this value at certain locations,
the values of Table~\ref{tab:fluxes} tell us that in the spatial average these energy 
losses cannot be balanced by the acoustic energy flux generated in our model. This
conclusion is emphasized by the fact that the values of Table~\ref{tab:fluxes} are
temporal maxima: the temporal mean would be lower. In order to be compatible with 
the observed quasi-steady Ca emission the injection of energy needs to be in the form 
of sustained short duration pulses as argued by \citep{hasan2008} but these pulses could 
probably not maintain the maximum values of acoustic flux as quoted in Table~\ref{tab:fluxes}.

Possibly with the exception of the case corresponding to the last row in 
Table~\ref{tab:fluxes}, the transverse excitation considered here rather 
correspond to the ``random walk phase'' of the model by \cite{cranmer2005}. 
Excitations corresponding to the ``jump phase'' with even higher velocity amplitudes 
than considered in the present paper might temporarily be capable of providing the required
energy flux. However, with a mean interval time of 360~s these jump events are probably 
not responsible for the heating observed in \ion{Ca}{ii} network grains, which requires a
more steady or high frequency source.

We have not considered photospheric radiative losses, which would considerably
damp the waves before they reach chromospheric heights \citep{carlsson+stein02}. 
If these radiative losses are 
taken into account, the fluxes would further be lowered. Also not all of the acoustic
energy flux would be available for radiative energy loss in the chromosphere depending
on the details of this NLTE process. All this implies that acoustic waves generated 
by transverse motions of the footpoints of magnetic network elements cannot
balance the chromospheric energy requirements of network regions.

This conclusion cannot be expected to drastically change when turning to three 
spatial dimensions. The details of the mode coupling and the partition of
energy fluxes to the various modes would become more complex but the share of energy 
that resides in the acoustic mode cannot be much larger than in the two-dimensional 
case. On the contrary, the energy flux generated at the footpoint of the magnetic 
element would have to be distributed to a larger area in three spatial dimension 
so that the spatial mean at $z=1000$~km would be lower.

We have only considered single, short duration, transverse pulses
for the wave excitation. A more realistic driver with sustained pulses of varying
lengths, velocities, and time intervals  would give
rise to highly non-linear dynamics, which might yield increased acoustic fluxes. 
Also we have not considered longitudinal wave excitation, which would be available 
primarily from global $p$-mode oscillations. The latter are expected to provide
low frequency slow mode waves to the outer atmosphere via magnetic portals in the 
presence of inclined strong magnetic fields, where they would be available for 
dissipation through shock formation 
\citep{michalitsanos1973,suematsu1990,hansteen2006,jefferies+al06}. In fact,
this mechanism would also work in the periphery of a vertically oriented flux tube,
where the field is strongly inclined with respect to the vertical direction.
Another source of energy that was not considered here may come from direct dissipation 
of magnetic fields through Ohmic dissipation.

\begin{acknowledgements}

We thank R.~Schlichenmaier for valuable comments on  the manuscript
and gratefully acknowledge the report by the anonymous referee, which 
helped substantially improve the presentation.
This work was supported by the German Academic Exchange
Service (DAAD), grant D/05/57687, and the Indian Department of
Science \& Technology (DST), grant DST/INT/DAAD/P146/2006.
\end{acknowledgements}

\bibliographystyle{aa}
\bibliography{12450}

\begin{thebibliography}{33}
\expandafter\ifx\csname natexlab\endcsname\relax\def\natexlab#1{#1}\fi

\bibitem[{{Berger} {et~al.}(2004){Berger}, {Rouppe van der Voort},
  {L{\"o}fdahl}, {Carlsson}, {Fossum}, {Hansteen}, {Marthinussen}, {Title}, \&
  {Scharmer}}]{berger2004}
{Berger}, T.~E., {Rouppe van der Voort}, L.~H.~M., {L{\"o}fdahl}, M.~G.,
  {et~al.} 2004, \aap, 428, 613

\bibitem[{{Berger} \& {Title}(1996)}]{berger1996}
{Berger}, T.~E. \& {Title}, A.~M. 1996, \apj, 463, 365

\bibitem[{{Berger} \& {Title}(2001)}]{berger2001}
{Berger}, T.~E. \& {Title}, A.~M. 2001, \apj, 553, 449

\bibitem[{{Bogdan} {et~al.}(2003){Bogdan}, {Carlsson}, {Hansteen}, {McMurry},
  {Rosenthal}, {Johnson}, {Petty-Powell}, {Zita}, {Stein}, {McIntosh}, \&
  {Nordlund}}]{bogdan2003}
{Bogdan}, T.~J., {Carlsson}, M., {Hansteen}, V.~H., {et~al.} 2003, \apj, 599,
  626

\bibitem[{{Bray} \& {Loughhead}(1974)}]{bray1974}
{Bray}, R.~J. \& {Loughhead}, R.~E. 1974, {The solar chromosphere} (The
  International Astrophysics Series, London: Chapman and Hall, 1974)

\bibitem[{{Cally}(2005)}]{cally2005}
{Cally}, P.~S. 2005, \mnras, 358, 353

\bibitem[{{Cally}(2007)}]{cally2007}
{Cally}, P.~S. 2007, Astronomische Nachrichten, 328, 286

\bibitem[{{Carlsson} \& {Stein}(2002)}]{carlsson+stein02}
{Carlsson}, M. \& {Stein}, R.~F. 2002, in ESA Special Publication, Vol. 505,
  SOLMAG 2002. Proceedings of the Magnetic Coupling of the Solar Atmosphere
  Euroconference, ed. H.~{Sawaya-Lacoste}, 293--300

\bibitem[{{Cranmer} \& {van Ballegooijen}(2005)}]{cranmer2005}
{Cranmer}, S.~R. \& {van Ballegooijen}, A.~A. 2005, \apjs, 156, 265

\bibitem[{{De Pontieu} {et~al.}(2007){De Pontieu}, {Hansteen}, {Rouppe van der
  Voort}, {van Noort}, \& {Carlsson}}]{pontieu2007}
{De Pontieu}, B., {Hansteen}, V.~H., {Rouppe van der Voort}, L., {van Noort},
  M., \& {Carlsson}, M. 2007, \apj, 655, 624

\bibitem[{{Devore}(1991)}]{devore1991}
{Devore}, C.~R. 1991, Journal of Computational Physics, 92, 142

\bibitem[{{Grossmann-Doerth} {et~al.}(1974){Grossmann-Doerth}, {Kneer}, \&
  {Uexk{\"u}ll}}]{grossmann1974}
{Grossmann-Doerth}, U., {Kneer}, F., \& {Uexk{\"u}ll}, M.~V. 1974, \solphys,
  37, 85

\bibitem[{{Hansteen} {et~al.}(2006){Hansteen}, {De Pontieu}, {Rouppe van der
  Voort}, {van Noort}, \& {Carlsson}}]{hansteen2006}
{Hansteen}, V.~H., {De Pontieu}, B., {Rouppe van der Voort}, L., {van Noort},
  M., \& {Carlsson}, M. 2006, \apjl, 647, L73

\bibitem[{{Hasan} {et~al.}(2000){Hasan}, {Kalkofen}, \& {van
  Ballegooijen}}]{hasan2000}
{Hasan}, S.~S., {Kalkofen}, W., \& {van Ballegooijen}, A.~A. 2000, \apjl, 535,
  L67

\bibitem[{{Hasan} \& {Ulmschneider}(2004)}]{hasan2004}
{Hasan}, S.~S. \& {Ulmschneider}, P. 2004, \aap, 422, 1085

\bibitem[{{Hasan} \& {van Ballegooijen}(2008)}]{hasan2008}
{Hasan}, S.~S. \& {van Ballegooijen}, A.~A. 2008, \apj, 680, 1542

\bibitem[{{Hasan} {et~al.}(2005){Hasan}, {van Ballegooijen}, {Kalkofen}, \&
  {Steiner}}]{hasan2005}
{Hasan}, S.~S., {van Ballegooijen}, A.~A., {Kalkofen}, W., \& {Steiner}, O.
  2005, \apj, 631, 1270

\bibitem[{{Jefferies} {et~al.}(2006){Jefferies}, {McIntosh}, {Armstrong},
  {Bogdan}, {Cacciani}, \& {Fleck}}]{jefferies+al06}
{Jefferies}, S.~M., {McIntosh}, S.~W., {Armstrong}, J.~D., {et~al.} 2006,
  \apjl, 648, L151

\bibitem[{{Khomenko} {et~al.}(2008){Khomenko}, {Collados}, \&
  {Felipe}}]{khomenko2008}
{Khomenko}, E., {Collados}, M., \& {Felipe}, T. 2008, \solphys, 251, 589

\bibitem[{{Lites} {et~al.}(1993){Lites}, {Rutten}, \& {Kalkofen}}]{lites1993}
{Lites}, B.~W., {Rutten}, R.~J., \& {Kalkofen}, W. 1993, \apj, 414, 345

\bibitem[{{Michalitsanos}(1973)}]{michalitsanos1973}
{Michalitsanos}, A.~G. 1973, \solphys, 30, 47

\bibitem[{{Muller}(1985)}]{muller1985}
{Muller}, R. 1985, \solphys, 100, 237

\bibitem[{{Muller} {et~al.}(1994){Muller}, {Roudier}, {Vigneau}, \&
  {Auffret}}]{muller1994}
{Muller}, R., {Roudier}, T., {Vigneau}, J., \& {Auffret}, H. 1994, \aap, 283,
  232

\bibitem[{{Nisenson} {et~al.}(2003){Nisenson}, {van Ballegooijen}, {de Wijn},
  \& {S\"utterlin}}]{nisenson+al03}
{Nisenson}, P., {van Ballegooijen}, A.~A., {de Wijn}, A.~G., \& {S\"utterlin},
  P. 2003, \apj, 587, 458

\bibitem[{{Oran} \& {Boris}(1987)}]{oran+boris1987}
{Oran}, E.~S. \& {Boris}, J.~P. 1987, {Numerical Simulation of Reactive Flow}
  (Elsevier)

\bibitem[{{Rosenthal} {et~al.}(2002){Rosenthal}, {Bogdan}, {Carlsson}, {Dorch},
  {Hansteen}, {McIntosh}, {McMurry}, {Nordlund}, \& {Stein}}]{rosenthal2002}
{Rosenthal}, C.~S., {Bogdan}, T.~J., {Carlsson}, M., {et~al.} 2002, \apj, 564,
  508

\bibitem[{{Schaffenberger} {et~al.}(2005){Schaffenberger},
  {Wedemeyer-B{\"o}hm}, {Steiner}, \& {Freytag}}]{schaffenberger2005}
{Schaffenberger}, W., {Wedemeyer-B{\"o}hm}, S., {Steiner}, O., \& {Freytag}, B.
  2005, in ESA Special Publication, Vol. 596, Chromospheric and Coronal
  Magnetic Fields, ed. D.~E. {Innes}, A.~{Lagg}, \& S.~A. {Solanki}

\bibitem[{{Sheminova} {et~al.}(2005){Sheminova}, {Rutten}, \& {Rouppe van der
  Voort}}]{sheminova2005}
{Sheminova}, V.~A., {Rutten}, R.~J., \& {Rouppe van der Voort}, L.~H.~M. 2005,
  \aap, 437, 1069

\bibitem[{{Solanki}(1993)}]{solanki1993}
{Solanki}, S.~K. 1993, Space Science Reviews, 63, 1

\bibitem[{{Steiner} {et~al.}(1994){Steiner}, {Kn{\"o}lker}, \&
  {Sch{\"u}ssler}}]{steiner1994}
{Steiner}, O., {Kn{\"o}lker}, M., \& {Sch{\"u}ssler}, M. 1994, in Solar Surface
  Magnetism, ed. R.~J. {Rutten} \& C.~J. {Schrijver}, 441--470

\bibitem[{{Steiner} {et~al.}(1986){Steiner}, {Pneuman}, \&
  {Stenflo}}]{steiner1986}
{Steiner}, O., {Pneuman}, G.~W., \& {Stenflo}, J.~O. 1986, \aap, 170, 126

\bibitem[{{Steiner} {et~al.}(2007){Steiner}, {Vigeesh}, {Krieger},
  {Wedemeyer-B{\"o}hm}, {Schaffenberger}, \& {Freytag}}]{steiner2007}
{Steiner}, O., {Vigeesh}, G., {Krieger}, L., {et~al.} 2007, Astronomische
  Nachrichten, 328, 323

\bibitem[{{Suematsu}(1990)}]{suematsu1990}
{Suematsu}, Y. 1990, in Lecture Notes in Physics, Berlin Springer Verlag, Vol.
  367, Progress of Seismology of the Sun and Stars, ed. Y.~{Osaki} \&
  H.~{Shibahashi}, 211--214

\end{thebibliography}

\end{document}